\documentclass[twocolumn,superscriptaddress,aps,showpacs,showkeys]{revtex4}
\usepackage{graphicx}
\usepackage{graphics}
\usepackage{dcolumn}
\usepackage{amssymb}
\usepackage{bm}
\usepackage{amsmath}
\usepackage{epsf,amsfonts}
\usepackage{epsfig}
\usepackage{indentfirst}
\usepackage{color}

\usepackage[utf8x]{inputenc}
\usepackage[english]{babel}
\usepackage[T1]{fontenc}

\DeclareMathOperator{\senh}{senh}
\DeclareMathOperator{\sen}{sen}

\newcommand{\be}{\begin{equation}}
\newcommand{\en}{\end{equation}}
\newcommand{\bea}{\begin{eqnarray}}
\newcommand{\ena}{\end{eqnarray}}
\newcommand{\ie}{\textsl{i.e.~}}

\def\1{1 \!\! 1}
\newcommand{\eqq}[1]{eq.({#1})}

\newcommand{\dd}{\textrm{d}}

\begin{document}
\title{Nonlocal Effects in Black Body Radiation}

\author{G.~N.~Bremm}
\affiliation{CBPF - Centro Brasileiro de Pesquisas F\'{\i}sicas \\
Xavier Sigaud st., 150, Urca,
CEP22290-180, Rio de Janeiro, Brazil}

\author{F.~T.~Falciano}
\affiliation{CBPF - Centro Brasileiro de Pesquisas F\'{\i}sicas \\
Xavier Sigaud st., 150, Urca,
CEP22290-180, Rio de Janeiro, Brazil}

\date{\today}

\begin{abstract}
Nonlocal electrodynamics is a formalism developed to include nonlocal effects in the measurement process caused by the non-inertial state of the observers. This theory modifies Maxwell's electrodynamics by eliminating the hypothesis of locality that assumes an accelerated observer simultaneously equivalent to a comoving inertial frame of reference. In this scenario, the transformation between an inertial and accelerated observer is generalized which affects the properties of physical fields. In particular, we analyze how an uniformly accelerated observer perceives a homogeneous and isotropic blackbody radiation. We show that all nonlocal effects are transient and most relevant in the first period of acceleration.
\end{abstract}



\maketitle

\section{Introduction}\label{intro}

In physics, the principle of relativity establishes the equivalence between all inertial observers but at the same time raises them to a special class in the sense that the laws of physics are the same in all inertial frames of references. The transition from classical mechanics to special relativity maintains this assumption intact and only modify the group of symmetry associated with these inertial observers. In classical mechanics we have the galilean invariance of Newtonian physics while in special relativity we have the Poincaré group connecting different inertial observers. To each of these groups of symmetry there is a geometrical absolute object and an invariant physical quantity associated to it. In particular, in classical mechanics the tridimensional euclidean metric is an absolute object and the length of material bodies is invariant under the action of the galilean group. Accordingly, in special relativity the absolute and invariant objects are respectively the minkowski four-dimensional metric, which in cartesian coordinates reads $\eta_{\mu\nu}=\mathrm{diag}(1,-1,-1,-1)$, and the spacetime interval defined by $\dd s^2=c^2 \dd t^2-\dd x^2-\dd y^2-\dd z^2$.

However, inertial frames of reference are only idealizations inasmuch real physical observers are always interacting and hence they are actually accelerated observers. In order to connect the laws of physics defined for inertial observers with actual measurements performed by accelerated observers there is an extra assumption that following  Bahram Mashhoon \cite{mashhoon1990}-\cite{mashhoon2010} we shall call the hypothesis of locality. This hypothesis states that an accelerated observer is instantaneously equivalent to a momentarily comoving inertial observer. In other words, the path of an accelerated observer can be understood as a continuous sequence of inertial observers with appropriate instantaneously velocities. This hypothesis of locality is consistent with the newtonian world-view of point-like particles where the state of a physical system is completely determined by the position and velocity of its parts at a given time. Notwithstanding, wave phenomena are intrinsic nonlocal and as Mashhoon have shown 
\cite{mashhoon1993}-\cite{mashhoon2007} this hypothesis of locality is, in general, only approximately valid. 

The accuracy of the locality approximation depends on the relative variability between the observer's velocity and the typical timescale of the system under consideration. Suppose that the physical process has a typical size $\lambda$ or a typical timescale that can always be associated with a length through the velocity of light, \ie $\lambda/c$ and let the magnitude of the observer's acceleration be $a$ such that the timescale over which his/her velocity changes be given by $c/a$ , or in terms of length $L= c^2/a$. The condition for the validity of the hypothesis of locality can be cast as 
\begin{equation} 
\lambda \ll L
\end{equation}

This relation encodes the idea that  during a measurement the velocity of the observer should not vary significantly such that he/she does not depart too much from an inertial frame of reference.

As an example, consider a monochromatic electromagnetic wave with frequency $\omega$. To properly measure the frequency of this wave, an observer needs to capture the oscillations of the electromagnetic fields.  The number of oscillations can vary with the adequacy of the experimental apparatus but he/she will need at minimum two oscillations for such a measurement. Thus, the experiment should last longer than the wave's period, \ie $2\pi/w$. If an observer has instantaneous velocity $\vec{v}(t)$, then the timescale over which it changes its velocity appreciably is $|\vec{v}|/|\vec{a}|$. Therefore the hypothesis of locality requires that $|\vec{v}|\gg |\vec{a}|/w$.

The two typical cases are for a linearly accelerated observer and for a rotating observer with fixed radius. For an observer describing a circle of radius r and angular velocity $\Omega$ the centripetal acceleration is given by $a_c=v^2/r=\Omega v$. Thus, in terms of the wave-length $\lambda =2\pi c/w$, for a linear acceleration $a_L$ the conditions for the hypothesis of locality reads $\lambda\ll c^2/a_L$. Similarly for a rotating observer we have $\lambda\ll c/\Omega$.

Generally, these quantities are too small to be detected in laboratory experiments since the Earth gravitational field gives ${c^2}/{g_{\bigoplus}} \simeq 1 \ \mathrm{light\ year} \simeq 10^{13}\ \mathrm{Km}$ while its rotation gives ${c}/{\Omega_{\bigoplus}} \simeq 28 \ \mathrm{a.\, u.} \simeq 5\times 10^{9}\ \mathrm{Km}$ that are much larger than typical dimensions of laboratory systems. Thus one should expect the hypothesis of locality to be very suitable to everyday physics. Notwithstanding, there are situations where it might break-down as for instance an electric charged particle interacting with an electromagnetic field. It is well known that charged particles irradiates when accelerated, hence, its equation of motion must include a term to account for its lost of energy. As a consequence the state of the accelerated particle is not completely specified by its position and velocity at a given instant of time, \ie the hypothesis of locality is violated in this case.

Let us consider an arbitrary physical field $\psi$ written in terms of a global inertial coordinate system $\mathbf{x}$. In an another inertial frame of reference $\mathbf{x}'$, associated with a moving observer, the same field becomes
\begin{equation}
\hat \psi (\mathbf{x}'(\tau))= \Lambda(\tau)\psi (\mathbf{x}(\tau))
\end{equation}
where $\Lambda$ is a Lorentz matrix connecting both systems and $\tau$ is the proper time of the observer. In the case of an accelerated observer, one shall use a set of vectors attached to him/her, namely his/her tetrad field, to project the $\psi$ field in his/her local frame of reference. Therefore we have
\begin{equation}
\hat \psi (\mathbf{x}(\tau)) = \Upsilon(\tau)\psi (\mathbf{x}(\tau)) \label{4.5.1}
\end{equation}
with the $\Upsilon$ matrix builded from the tetrad field.

Let us designate by $\hat \varPsi (\tau)$ the actual measurement performed by the observer. Then the hypothesis of locality identifies $\hat \varPsi (\tau) = \hat \psi(\tau)$, \ie the observer measures exactly the instantaneously projected field $\hat \psi(\tau)$.

In order to account for nonlocal effects due to acceleration, one has to generalize this relation. Following Mashhoon's ansatz \cite{mashhoon1993}, we shall maintain a linear relation between the physical field $\hat \psi(\tau)$ and the measured field $\hat \varPsi (\tau)$. The most general linear relation that satisfies causality is of the form
\begin{equation}
\hat \varPsi (\tau) = \hat \psi(\tau) + \int_{\tau_0}^{\tau} K(\tau,\tau')\hat \psi(\tau')\rm{d}\tau', \label{428}
\end{equation}
with $\tau_0$ being the moment when acceleration starts and $K(\tau,\tau')$ is the \textit{kernel} associated with the observer's acceleration. In particular, without acceleration, the kernel must vanish so that we recover the relation $\hat \varPsi (\tau) = \hat \psi(\tau)$.

The \textit{ansatz} \eqq{\ref{428}} is a Volterra integral equation of the second kind which, for a given kernel, uniquely determine the field $\hat \varPsi (\tau)$ in terms of $\hat \psi (\tau')$ (see ref.s \cite{tricomi1957}-\cite{weber2005}). The choice of the kernel can be motivated by requiring that no electromagnetic radiation field can be at rest with respect to any observer, inertial or accelerated. In other words, if $ \hat \psi(\tau)$ is a static field for a given observer than necessarily $\hat \varPsi (\tau)$ is also static. This condition implies that
\begin{equation}
\Upsilon(\tau) + \int_{\tau_0}^{\tau} K(\tau,\tau')\Upsilon(\tau') \rm{d}\tau' \equiv \Upsilon_{0}\ , \label{430}
\end{equation}
where $\Upsilon_{0}$ is a constant. This relation still doesn't determine uniquely the kernel so we must add the assumption that the kernel is a function of a single variable. 

There are two proposals in the literature for single variable kernels (see ref.'s \cite{chicone2002}-\cite{chicone2002b}), namely, the kinetic kernel $K(\tau,\tau') = k(\tau')$ and the dynamic kernel $K(\tau,\tau') = k(\tau - \tau') $. However, the dynamic kernel might endure even after the end of the acceleration hence producing, in some cases, divergencies of the fields. For this reason, we shall hereon focus only on the kinetic kernel.

Differentiating \eqq{\ref{430}} we find
\begin{equation}
k(\tau)=-\frac{\rm{d}\Upsilon(\tau)}{\rm{d}\tau}\Upsilon^{-1}(\tau)\ , \label{kernel}
\end{equation}
where the existence of the inverse matrix $\Upsilon^{-1}(\tau)$ is guaranteed by the existence of the inverse of the tetrad field. Note that as soon as the acceleration stops the $\Upsilon(\tau)$ matrix no longer varies and the kinetic kernel vanishes. This shows that the kinetic kernel is free of the endurance issue of the dynamic kernel. Using the above result we have
\begin{equation}\label{nonlocalfields1}
\hat \varPsi (\tau) = \hat \psi(\tau) - \int_{\tau_0}^{\tau} \frac{\rm{d}\Upsilon(\tau')}{\rm{d}\tau'}\Upsilon^{-1}(\tau') \hat \psi(\tau')\rm{d}\tau'\ ,
\end{equation}
or integrating by parts
\begin{equation}\label{nonlocalfields1a}
\hat \varPsi (\tau) = \hat \psi(\tau_0) + \int_{\tau_0}^{\tau} \Upsilon(\tau') \frac{\rm{d}\psi(\tau')}{\rm{d}\tau'}\dd \tau'\ .
\end{equation}

One can immediately check from \eqq{\ref{nonlocalfields1a}} that two generic observers will always agree if the physical field is constant or not. Indeed, if an observer measures a constant $\hat \psi(\tau)$ field then the other observer will also measure $\hat \varPsi (\tau) = \hat \psi(\tau_0)$.

In this paper we are interested in examining the acceleration induced nonlocal effects in a black body radiation. As it is well know, the universe is filled with a homogenous and isotropic radiation thermal bath that presents the most perfect black body spectrum ever measured. Thus, it is suitable to analyze nonlocal contribution to this radiation field. The paper is organized as follows. In the next section we apply the nonlocal theory for electromagnetic fields and construct the nonlocal energy-momentum tensor measured by an accelerated observer. In section~\ref{sec:HIBBR} we describe the black body radiation field and the average procedure to achieve a homogenous and isotropic radiation field. Section~\ref{sec:TheoBg} we analyze the nonlocal effects and conclude with some final remarks.

\section{Nonlocal Electrodynamics}\label{NonlocalEletro}

The nonlocal formalism described in the last section is general in the sense that $\psi(\tau)$ can be any physical field (see ref.'s \cite{chicone2007}-\cite{mashhoon1986}). In particular, for an electromagnetic field, the Faraday tensor has two spacetime indices. Given the tensor field $F_{\mu\nu}$, an accelerated observer will measure the projected tensor
\begin{equation}
F'^{(a)(b)} =  {e^{(a)}}_{\mu}{e^{(b)}}_{\nu}F^{\mu\nu},
\end{equation}
where ${e^{(a)}}_{\mu}$ is its associated tetrad field. In what follows, it will be convenient to define a six-dimensional vector
\[ \mathbf{F} \equiv \left(\begin{array}{c}
\mathbf{E} \\
c \mathbf{B} \end{array} \right) \]
such that 
\begin{equation}\label{nonlocaleletromagfields}
\mathbf{F'}(\tau) = \mathbf{\Upsilon}(\tau) \mathbf{F}(\tau) \ ,
\end{equation}
where $\mathbf{\Upsilon}$ is a $6\times 6$ matrix.  The six-dimensional vector $\mathbf{F}(\tau)$ plays the role of the $\psi(\tau)$ field, hence, it is the electromagnetic field measured by an inertial observer. The hypothesis of locality claims that the accelerated observer will measure $\mathbf{F'}(\tau)$ given by \eqq{\ref{nonlocaleletromagfields}}.
However, accordingly to  \eqq{\ref{nonlocalfields1}}, the nonlocal electromagnetic fields $\mathbf{\mathcal{F}}=(\mathbf{\mathcal{E}},c\mathbf{\mathcal{B}})$ are given by 
\begin{equation}\label{nonlocaleletromagfields2}
 \left(\begin{array}{c}
\mathbf{\mathcal{E}} \\
c\mathbf{\mathcal{B}} \end{array} \right)
= \left(\begin{array}{c}
\mathbf{E'} \\
c\mathbf{B'} \end{array} \right)  - \int_{0}^{\tau} \frac{\mathrm{d}\mathbf{\Upsilon}}{\mathrm{d}\tau'}\left(\begin{array}{c}
\mathbf{E} \\
c\mathbf{B} \end{array} \right) \mathrm{d}\tau'\quad .
\end{equation}

The nonlocal fields $(\mathbf{\mathcal{E}},c\mathbf{\mathcal{B}})$ depend on the observer's world-line. Thus, to go further on our analysis, we must specify a particular trajectory (for a general discussion see \cite{maluf2010}). We shall develop our analysis for a linear accelerated observer. Since we are neglecting any gravitational effects, in other words, the background is the Minkowski flat spacetime, we can choose, without restriction, the observer trajectory along the $\hat{z}$ direction.

\subsection{Linear Accelerated Observer}\label{LinearAccObs}

Let us consider a linearly accelerated observer along a given direction, say the $\hat{z}$ axis. If its comoving acceleration is a constant $g_0$ then the Lorentz transformations give
\begin{equation}\label{acrep}
a = \left(1 - \frac{v^2}{c^2}\right)^{\frac{3}{2}}g_0 \quad ,
\end{equation}
where $a$ is the observe's acceleration along the $\hat{z}$ direction. Integrating \eqq{\ref{acrep}} we find the well known hyperbolic trajectory for a rindler observer (see ref.'s~\cite{dinverno}-\cite{gravitation})
\begin{equation}
\left(z - z_0 + \frac{c^2}{g_0}\right)^{2} - (ct)^2 = \frac{c^4}{{g_0}^2}\quad .
\end{equation}

Using the observer's proper time $\mathrm{d} \tau = \sqrt{1 - {v^2}/{c^2}}\  \dd t$, we can parametrize the hyperbolic motion as 
\begin{align}\label{tacel}
t &= \frac{c}{g_0}\senh \theta\quad , \\
z & = z_0 + \frac{c^2}{g_0}\left(\cosh\theta-1\right)\nonumber
\end{align}
where we have define $\theta(\tau) \equiv \frac{g_0}{c}\tau$ for later convenience. The perpendicular directions remain intact, \ie $x = x_0$ and $y = y_0$. The tetrad fields associated with this accelerated observer read
\begin{displaymath}
{e_{(a)}}^{\mu} = \left(
\begin{array}{cccc}
\cosh \theta & 0 & 0 & \senh \theta \\
0 & 1 & 0 & 0 \\
0 & 0 & 1 & 0 \\
\senh \theta & 0 & 0 & \cosh \theta \\
\end{array} \right)\ .
\end{displaymath}

A direct calculation shows that the electric and magnetic fields $\left(\mathbf{E}' , c \mathbf{B}'\right)$ are given by 
\begin{align*}
{E'}^{i} &=E^j \left({e_{(0)}}^{0}{e_{(i)}}^{j} - {e_{(0)}}^{j}{e_{(i)}}^{0}\right)- c B^m\epsilon_{jlm}{e_{(0)}}^{j}{e_{(i)}}^{l}\ ,\\
{B'}^i &= \epsilon^{ijk} \left(\frac{1}{2}{e_{(j)}}^{l}{e_{(k)}}^{n}\epsilon_{lnp} B^{p} - {e_{(j)}}^{0}{e_{(k)}}^{l}\frac{E_l}{c} \right)\ ,
\end{align*}
or explicitly in components, the local electromagnetic fields $\left(\mathbf{E}' , c \mathbf{B}'\right)$ can be written in terms of the background fields as 
\begin{equation}
\begin{split}
& {E'}_{1} = {E}_{1} \cosh \theta - c B_{2} \senh \theta\quad , \\
& {E'}_{2} = {E}_{2} \cosh \theta + c B_{1} \senh \theta\quad  , \\  \label{Eacel}
& {E'}_{3} = {E}_{3}\quad ,
\end{split} 
\end{equation}
and
\begin{equation}
\begin{split}
& {B'}_{1} = {B}_{1} \cosh \theta + \frac{E_2}{c} \senh \theta\quad  , \\
& {B'}_{2} = {B}_{2} \cosh \theta - \frac{E_1}{c} \senh \theta\quad , \\ \label{Bacel}
& {B'}_{3} = {B}_{3}\quad .
\end{split}
\end{equation}

The above equation allow us to identify the six by six $\mathbf{\Upsilon}(\tau)$ matrix appearing in \eqq{\ref{nonlocaleletromagfields}} as
\[
\mathbf{\Upsilon} = \left(
\begin{array}{cc}
C & S\\
-S & C
\end{array}\right)
\]
where $C$ and $S$ are two three by three matrix given by
\[ C = \left(
\begin{array}{ccc}
\cosh\theta & 0 & 0 \\
0 & \cosh\theta & 0 \\
0 & 0 & 1 \\
\end{array} \right) \ \ , \ \ 
S = \left(
\begin{array}{ccc}
 0 & -\senh\theta & 0 \\
 \senh\theta & 0& 0 \\
0 & 0 & 0 \\
\end{array} \right)\ .
\]

The nonlocal fields are obtained by using $\mathbf{\Upsilon}(\tau)$ in \eqq{\ref{nonlocaleletromagfields2}}, \ie
\begin{equation}
\begin{split}
& {\mathcal{E}_1}(\tau) = {E'}_{1}(\tau) + c \frac{g_{0}}{c}\int_{0}^{\tau} {B'}_{2}(\tau')\rm{d}\tau'\ , \\
& {\mathcal{E}_2}(\tau) = {E'}_{2}(\tau) - c \frac{g_{0}}{c}\int_{0}^{\tau}{B'}_{1}(\tau')\rm{d}\tau' \ ,\\ 
& {\mathcal{E}_3}(\tau) = {E'}_{3}(\tau)\ ,
\end{split}
\end{equation}
and 
\begin{equation}
\begin{split}
& {\mathcal{B}_1}(\tau) = {B'}_{1}(\tau)  - \frac{g_{0}}{c}\int_{0}^{\tau}\frac{{E'}_{2}(\tau')}{c}\rm{d}\tau' \ , \\
& {\mathcal{B}_2}(\tau) = {B'}_{2}(\tau)  + \frac{g_{0}}{c}\int_{0}^{\tau}\frac{{E'}_{1}(\tau')}{c}\rm{d}\tau'\ , \\
& {\mathcal{B}_3}(\tau) = {B'}_{3}(\tau)\ .
\end{split}
\end{equation}

The thermal properties of the electromagnetic radiation fields are encoded in the decomposition of the energy-momentum tensor. This decomposition depends explicitly on the observer's world-line and hence will also carry nonlocal effects. For an arbitrary observer, the local energy-momentum tensor is simply the projection of the standard energy-momentum tensor in its tetrad field, \ie
\begin{align} \label{E-M'}
{T'}_{ab} &= {e_{(a)}}^{\mu}{e_{(b)}}^{\nu} {T}_{\mu\nu} \\
&= -\frac{1}{\mu_0}\left({F'}_{ac}{{F'}_{b}}^{c} - \frac{1}{4}\eta_{ab}{F'}^{cd}{F'}_{cd} \right)\ .\nonumber
\end{align}

Projecting the electromagnetic energy-momentum tensor along and perpendicular to the observer's world-line, we can define the thermodynamics quantities such as  the energy density, isotropic pressure, Poynting vector and Maxwell's stress tensor. 

Let the observer's world-line be given by the velocity field $v^\mu$. The energy density $\rho$ is defined as the double projection of $T_{\mu\nu}$ along the observer's world-line, while the isotropic pressure $p$ is one-third of the energy density minus its trace. The Poynting vector $\vec{S}$ is given by projecting one indices of the $T_{\mu\nu}$ along the observer's trajectory and the other in its local space by using the projector $h^{\mu\nu}=\eta^{\mu\nu}-v^{\mu}v^{\nu}$. Finally, the Maxwell's stress tensor $T_{ij}$ is defined as the double projection in the observer's local space. Thus, we have
\begin{align}\label{decompTmunu}
\rho'&= \frac{\epsilon_0}{2}\left(\mathbf{E'}^2 + c^2 \mathbf{B'}^2\right)\qquad , \qquad \quad p'=\frac13 \rho' \nonumber\\
{S'}^{i} &= \frac{1}{\mu_0}\left( \mathbf{E'} \times \mathbf{B'}\right)^i\quad  \quad \quad \ , \\
{T'}_{ij} &=\epsilon_0 \left[  \frac{1}{2}\left(\mathbf{E'}^2 + c^2 \mathbf{B'}^2\right)\delta_{ij} -({E'}_i{E'}_j + c^2{B'}_i{B'}_j)  \right]\ .\nonumber
\end{align}

The transformation in the electromagnetic fields, \eqq{\ref{Eacel}}-(\ref{Bacel}), induces a transformation in the energy-momentum tensor such that
\begin{align*}
&\left(\begin{array}{c} T'_{10}\\  T'_{13} \end{array}\right) 
= \mathcal{R}(\theta)
\left( \begin{array}{c} T_{10}\\ T_{13} \end{array}\right) \\
&\left(\begin{array}{c} T'_{20}\\  T'_{23} \end{array}\right) 
=\mathcal{R}(\theta)
\left( \begin{array}{c} T_{20}\\ T_{23} \end{array}\right) \\
&\left(\begin{array}{cc} T'_{11}& T'_{12}\\  T'_{21}& T'_{22} \end{array}\right) 
= \left(\begin{array}{cc} T_{11}& T_{12}\\  T_{21}& T_{22} \end{array}\right) \\
&\left(\begin{array}{cc} T'_{00}& T'_{03}\\  T'_{30}& T'_{33} \end{array}\right) 
=\mathcal{R}^T(\theta)
\left(\begin{array}{cc} T_{00}& T_{03}\\  T_{30}& T_{33} \end{array}\right) 
\mathcal{R}(\theta)\\
\end{align*}with the two by two matrix given by 
\begin{align*}
\mathcal{R}(\theta)= \left( \begin{array}{cc} \cosh \theta & \sinh \theta\\ \sinh \theta & \cosh \theta \end{array}\right)
\end{align*}

Therefore, the description of an electromagnetic radiation in terms of the thermodynamics quantities \eqq{\ref{decompTmunu}} depends on the state of motion of the observer. We shall be interested in how nonlocal effects change these properties. In particular we shall analyze the case for a homogeneous and isotropic black body radiation.

\section{Homogeneous and Isotropic Black Body Radiation}\label{sec:HIBBR}

As it is well known, a black body radiation is a thermal radiation whose spectrum has an universal feature, \ie its spectral distribution satisfies Planck's law and is completely characterized by its temperature. Let $\rho_T(\nu)\mathrm{d}\nu$ be the energy density contained in the range of frequencies $\nu$ and $\nu + \dd \nu$.

The black body Planck distribution is given by
\begin{equation}
\rho (\nu) \mathrm{d}\nu  = \frac{8 \pi h}{c^3}\frac{\nu^3}{e^{\beta h \nu} - 1}\mathrm{d}\nu \quad , \label{leiplanck}
\end{equation}
with $\beta^{-1}=k_BT$, $k_B$ is the Boltzmann constant and $T$ the temperature. Integrating over all frequencies we obtain the Stefan-Boltzmann law 
\begin{equation}
\rho = \frac{4 \sigma}{c} T^4\quad , \label{stefan}
\end{equation}
where $\sigma \equiv \frac{2{k_{B}}^{4}\pi^{5}}{15c^2h^3} \approx 5,67 \times 10^{-8} \ \mathrm{J \cdot s^{-1} \cdot m^{-2} \cdot K^{-4}}$.

Given a generic bath of radiation, the energy density depends on both the position and on the time, \ie $\rho=\rho(\mathbf{r},t)$. However, a homogenous and isotropic radiation must be such that the average of the electric and magnetic fields vanish. In this case, the average energy density has no spatial dependence and becomes only a function of time.

Consider an ensemble of identical systems and let us define the ensemble averaged value of a quantity $\Sigma(\mathbf{r},t)$ by
\begin{equation}
\langle  \Sigma(\mathbf{r},t) \rangle \equiv \lim_{N\rightarrow \infty} \frac{1}{N} \sum_{i=1}^{N}\Sigma_{i}(\mathbf{r},t)\quad .
\end{equation}

It is evident that if $\Sigma(\mathbf{r},t)$ is an incoherent quantity then its average will not depend on the position, \ie $\langle  \Sigma(\mathbf{r},t) \rangle =\Sigma(t)$. An incoherent electromagnetic field (ref.'s~\cite{bourret1960}-\cite{tolman1934}) must have $\langle  \mathbf{E}(\mathbf{r},t) \rangle=\langle  \mathbf{B}(\mathbf{r},t) \rangle=0$. On the other hand, its average energy density depends on the square of the fields
\begin{equation}
\rho= \langle \rho(\mathbf{r},t)\rangle = \frac{\epsilon_{0}}{2} \left( \sum_{i=1}^{3} \langle {E_i}^2\rangle +c^2 \sum_{i=1}^{3} \langle {B_i}^2\rangle\right)\ .
\end{equation}
For an incoherent field we expect to have
\begin{equation}
\langle {E_i}^2\rangle = c^2 \langle {B_i}^2\rangle = \frac{1}{3\epsilon_{0}} \rho\quad .
\end{equation}

A third condition for homogeneity and isotropy is that the electromagnetic fields have no energy flux, hence the fields must satisfy 
\begin{equation}
\langle E_{i}(\mathbf{r}_{1},t_{1})B_{j}(\mathbf{r}_{2},t_{2})\rangle = 0\ , \label{descoerencia}
\end{equation}
in such a way that it has zero Poynting vector. These conditions can be put in a more compact expression as
\begin{equation}\label{503}
\langle  E_{i}(\mathbf{r},t)E_{j}(\mathbf{r},t) \rangle =
\langle  B_{i}(\mathbf{r},t)B_{j}(\mathbf{r},t) \rangle = \frac{\rho}{3\epsilon_{0}c^2}\delta_{ij}\ .
\end{equation}

Even though in general the average quantities can depend on time, we will assume hereon that the average thermodynamics quantities are constant.

In order to describe the correlation of the same components of the fields but in two different positions and/or times we shall assume that the fields are stationary in space and time such that its correlation depends only on the differences $\mathbf{r} = \mathbf{r_2} - \mathbf{r_1}$ and $t = t_2 - t_1$. In this way we can define a coherent function as
\begin{align}
\mathcal{C}_{ij}(\mathbf{r}, t) &\equiv \langle E_{i}(\mathbf{r}_{1}, t_{1}) E_{j}(\mathbf{r}_{2}, t_{2})\rangle \\
&= c^2 \langle B_{i}(\mathbf{r}_{1}, t_{1}) B_{j}(\mathbf{r}_{2}, t_{2})\rangle\ .\nonumber
\end{align}

In vacuum, the electromagnetic fields satisfy the wave equation which implies that the above function must also satisfy an identical wave equation
\begin{equation}
\left(\nabla^2 - \dfrac{1}{c^2}\partial_t^{2}\right)\mathcal{C}_{ij}(\mathbf{r}, t) = 0\ .
\end{equation}

Then, it follows that $\mathcal{C}_{ij}$ can be written as a linear combination of periodic functions
\begin{equation} \label{57}
\mathcal{C}_{ij}(\mathbf{r}, t) = \int f_{ij}(\mathbf{k})\cos{(kct)}\exp\left({i\mathbf{k}\cdot\mathbf{r}}\right) \mathrm{d}^3k\quad \ ,
\end{equation}
with $f_{ij}(\mathbf{k}) = f_{ij}(\mathbf{-k}) $ .

There is a close analogy between the present situation and the hydrodynamic flow of a homogeneous fluid (see ref.\cite{batchelor1959}). In particular, the vanishing of the divergence of the electric field, ${\nabla}. \mathbf{E}=0$, is analogous to the vanishing of the divergence of the velocity field for an incompressible fluid ${\nabla}.\mathbf{v}=0$. In this case we have
\begin{equation}
f_{ij}(\mathbf{k}) = A(k)k_ik_j + B(k) \delta_{ij}\quad ,
\end{equation}
where in principle $A(k)$ and $B(k)$ are arbitrary real functions. Notwithstanding, the continuity condition 
\begin{equation}
k^if_{ij}(\mathbf{k}) = k^jf_{ij}(\mathbf{k}) = 0\ ,
\end{equation}
implies that $B(k) = - A(k)k^2$. Therefore, the function $f_{ij}(\mathbf{k})$ depends only on one generic function and can be written as
\begin{equation}\label{58} 
f_{ij}(\mathbf{k}) = A(k)(k^2 \delta_{ij} - k_i k_j)\quad . 
\end{equation}

Using this result back in \eqq{\ref{57}}, the coherence function for the same spatial point $\mathbf{r} = 0$ becomes 
\begin{align} \label{59}
&\mathcal{C}_{ij}(t) =\\
&=\int_0^\infty\mathrm{d}k  \int_0^{\pi}\mathrm{d}\theta \sen{\theta} \int_0^{2\pi}\mathrm{d}\phi \, A(k)k^2(k^2 \delta_{ij} - k_{i}k_{j}) \cos{(kct)} 
\nonumber
\end{align}
To simplify the above integral we recall that 
\begin{equation}
\int_0^{2\pi}\int_0^\pi k_ik_j \sen{\theta}\mathrm{d}\theta\mathrm{d}\phi=\frac{4\pi}{3}k^2\delta_{ij}\quad ,
\end{equation}
which give us 
\begin{equation}\label{60}
\mathcal{C}_{ij}(t) = \dfrac{8\pi}{3}\delta_{ij}\int_0^\infty A(k) k^4\cos{(kct)} \mathrm{d}k\quad . 
\end{equation}

For a black body radiation 
\begin{equation}
k A(k) = \frac{\hbar c}{8\pi^3\epsilon_{0}}\frac{1}{\exp(\hbar \beta ck) - 1}
\end{equation}
and hence
\begin{equation}
\mathcal{C}_{ij}(t) = \frac{\hbar c}{3\pi^2\epsilon_{0}}\delta_{ij}\int_0^\infty\frac{k^{3}}{\exp(\hbar \beta c k) - 1}\cos(kct)\mathrm{d}k  \quad.
\end{equation}

Defining the quantity $\xi \equiv \frac{\pi }{\hbar \beta}t$, direct integration gives 
\begin{equation}
\mathcal{C}_{ij}(t) = - \frac{ \pi^2}{6 \epsilon_{0}\hbar^3c^3\beta^4 }\mathcal{\mathcal{L}}'''(\xi)\delta_{ij}, \label{fcor}
\end{equation}
where $\mathcal{\mathcal{L}}(x)$ is the Langevin function defined as
\begin{equation}
\mathcal{\mathcal{L}}(x) \equiv \coth x - \frac{1}{x}\quad .
\end{equation}

\begin{figure}[!h]
\centering
\includegraphics[width=8.5cm]{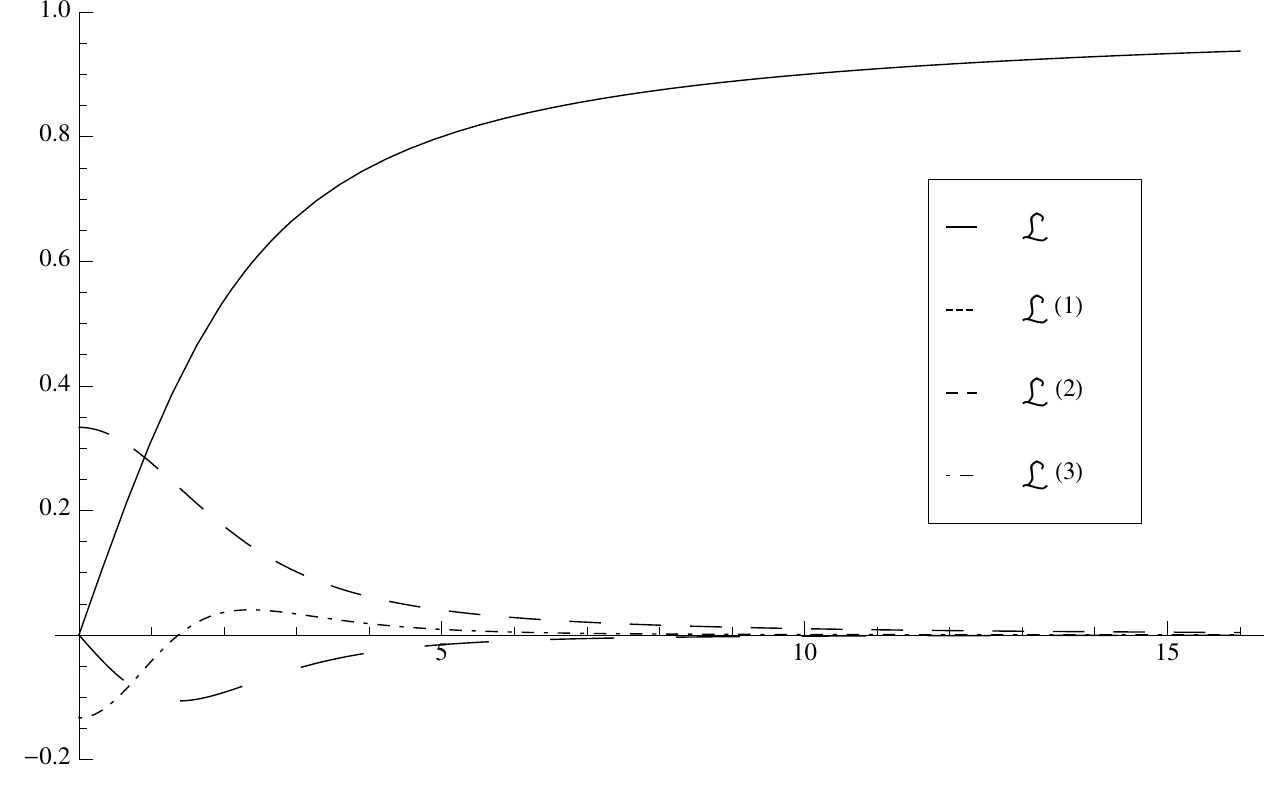}
\caption{\label{graficos_L}Langevin function and its first three derivatives.}
\end{figure}

The general behavior of the $\mathcal{\mathcal{L}}(x)$ function and its three first derivatives is plotted in figure \ref{graficos_L}. Note that these functions are smooth and restricted to the interval $\left[-0.2,1\right]$. In particular, $\mathcal{L}(x \rightarrow \infty) = 1$ while its derivatives go to zero for the same limit. Thus, the correlation given by \eqq{\ref{fcor}} decays as time differences increases. Another property worth mentioning is that the Langevin function has a symmetry given by 
\begin{equation}
\mathcal{\mathcal{L}}^{(i)}(-x) = (-1)^{i+1}\mathcal{\mathcal{L}}^{(i)}(x)\quad \mbox{for} \quad  i = 0,1,2,\dots
\end{equation}

\section{Black Body Radiation in an Accelerated Frame}\label{sec:TheoBg}

In the previous sections, we have developed the mathematical machinery to describe the nonlocal effects in a thermal bath. However, the thermodynamics properties of radiation fields depend on the observer's state of motion even assuming the hypothesis of locality. An observer moving through an homogenous and isotropic thermal bath will, in general, detect a non-zero Poynting vector even though an inertial observer at rest with respect to the same radiation field will measure zero Poynting vector. In order to extract the nonlocal effects we have first to disentangle it from the common local relativistic effects.

The nonlocal effects are taken into account by the map $(\mathbf{E}',\mathbf{B}') \rightarrow (\mathbf{\mathcal{E}},\mathbf{\mathcal{B}})$ in the energy-momentum tensor \eqq{\ref{E-M'}}
\begin{equation}
\mathcal{T}_{\mu\nu} = -\frac{1}{\mu_0}\left(\mathcal{F}_{\mu\alpha}{\mathcal{F}_{\nu}}^{\alpha} - \frac{1}{4}\eta_{\mu\nu}\mathcal{F}^{\alpha\beta}\mathcal{F}_{\alpha\beta} \right)\ ,
\end{equation}
where $\mathcal{F}_{\alpha\beta}$ is the nonlocal Faraday tensor. The average value of each of its components can be calculated by using the properties of homogeneity and isotropy introduced in section~\ref{sec:HIBBR}.

\subsection{Energy Density}

The average energy density measured by an accelerated observer is given by
\begin{equation} \label{densacel}
\rho_{ac} = \frac{\epsilon_{0}}{2} \sum_{i=1}^{3} \left( \langle {{\mathcal{E}}_i}^2\rangle +  c^2 \langle {{\mathcal{B}}_i}^2\rangle \right)\quad.
\end{equation}
A typical term of this expression is 
\[
\langle {\mathcal{E}_1}{\mathcal{E}_1}\rangle = \langle {\mathcal{E}_1}{\mathcal{E}_1}\rangle _{loc}+\langle {\mathcal{E}_1}{\mathcal{E}_1}\rangle_{nl1}+\langle {\mathcal{E}_1}{\mathcal{E}_1}\rangle _{nl2}\qquad \\
\]
where the subscripts stand for the nature of each term. The first term with subscript ``${loc}$'' is simply the common local term, ``${nl1}$'' is the first nonlocal corretion that is linear in the observer's acceleration and the last one is quadratic. The relativistic local part is given by
\[
\langle {\mathcal{E}_1}{\mathcal{E}_1}\rangle _{loc}=\frac{\rho}{3\epsilon_{0}}(1+2 \senh^2 \theta) 
\]

The ``${nl1}$'' has integral terms of $\langle E_{i}(\tau')E_{j}(\tau)\rangle$ and $\langle B_{i}(\tau')B_{j}(\tau)\rangle$. These quantities can be associated to the correlation function \eqq{\ref{fcor}} for different proper times such that
\begin{equation}
\langle E_{i}(\tau')E_{j}(\tau)\rangle  = \mathcal{C}_{ij}\left(t'(\tau') - t(\tau) \right) = \mathcal{C}(\tau', \tau)\delta_{ij}\quad ,
\end{equation}
with
\begin{equation}\label{corre}
\mathcal{C}(\tau', \tau) \equiv - \frac{ \pi^2}{6 \epsilon_{0}\hbar^3c^3\beta^4 }\mathcal{\mathcal{L}}'''\left(\frac{\pi }{\hbar\beta}\frac{c}{g_0}\Big( \senh \theta' - \senh \theta\Big)\right)\ .
\end{equation}

We can then write 
\begin{align}\label{EEnl1}
\langle {\mathcal{E}_1}{\mathcal{E}_1}\rangle_{nl1}=& -2\frac{g_{0}}{c} \left( \cosh{\theta}\int_{0}^{\tau}\mathcal{C}(\tau',\tau)\senh \theta'\rm{d}\tau' \right.\nonumber\\
& \ \left.+ \senh{\theta}\int_{0}^{\tau}\mathcal{C}(\tau',\tau)\cosh \theta'\rm{d}\tau' \right) \, .
\end{align}

Simirlarly, the quadratic term reads
\begin{align}\label{nl2}
\langle {\mathcal{E}_1}{\mathcal{E}_1}\rangle_{nl2}=&
\left(\frac{g_{0}}{c}\right)^2\int_{0}^{\tau}{\mathrm{d}\tau'}\int_{0}^{\tau}{\mathrm{d}\tau''} \mathcal{C}(\tau'',\tau') \times \\
 &\times \Big(\senh \theta' \senh \theta''+\cosh \theta'\cosh \theta''\Big) \ . \nonumber 
\end{align}

The other terms present in \eqq{\ref{densacel}} are trivial or equals the above result. A straightforward calculation shows that
\begin{align}\label{relacoes1}
\langle \mathcal{E}_2\mathcal{E}_2 \rangle =c^2 \langle \mathcal{B}_1\mathcal{B}_1 \rangle
=c^2 \langle \mathcal{B}_2\mathcal{B}_2 \rangle 
=\langle \mathcal{E}_1\mathcal{E}_1 \rangle
\end{align}
and 
\begin{align}\label{relacoes2}
\langle \mathcal{E}_3\mathcal{E}_3 \rangle =c^2 \langle \mathcal{B}_3\mathcal{B}_3 \rangle
=\frac{\rho}{3\epsilon_{0}}
\end{align}

The nonlocal effects appear as power of ${g_{0}}/{c}$ which is expected to be small. Thus, if the integrals in \eqq{\ref{nl2}} do not diverge, we can neglect the second order correction and keep just the linear term. To evaluate these integrals let us define
\begin{align}
\mathrm{I}_{CC} (\tau) &\equiv \int_{0}^{\tau}\int_{0}^{\tau} \mathcal{C}(\tau'',\tau')\cosh \theta'\cosh \theta'' \mathrm{d}\tau'\mathrm{d}\tau''\ ,\\
\mathrm{I}_{SS} (\tau) &\equiv \int_{0}^{\tau}\int_{0}^{\tau} \mathcal{C}(\tau'',\tau')\senh \theta'\senh \theta'' \mathrm{d}\tau'\mathrm{d}\tau''\ .
\end{align}

The first integral can be directly integrated to give
\begin{equation}\label{ICCeval}
\mathrm{I}_{CC} (\tau) = \frac{1}{3 \epsilon_{0}c^3\hbar \beta^2 }\left( \mathcal{L}'(0) - \mathcal{L}'\left( \frac{\pi}{\hbar \beta}\frac{c}{g_0} \senh \theta \right) \right)\ ,
\end{equation}
where we have used the property $\mathcal{L}'(-x) = \mathcal{L}'(x)$. Note that only the constant term survives for long times since $\lim_{\, x \rightarrow \infty}\mathcal{L}'(x)=0$. The other integral can be recast as 
\begin{equation}\label{ISS}
\mathrm{I}_{SS} (\tau) =- \frac{1}{6 \epsilon_{0}c^3\hbar \beta^2 }\int_{0}^{x}\int_{0}^{x} \frac{\mathcal{L}'''\left( x'' - x' \right)x''x'}{\sqrt{\alpha^2 + {x''}^2}\sqrt{\alpha^2 + {x'}^2}} \mathrm{d}x''\mathrm{d}x'\quad ,
\end{equation}
where $\alpha \equiv \frac{\pi }{\hbar \beta}\frac{c}{g_0}$ is a dimensionless parameter. There is no analytical solution for this integral so we shall approximate
\begin{equation}\label{approx}
\frac{x}{\sqrt{\alpha^2 + x^2}} \cong \left\{
\begin{array}{rc}
\frac{x}{\alpha} &\mbox{for}\quad x \le \alpha\\
\\
1 &\mbox{for}\quad x>\alpha
\end{array}\right.
\end{equation}

\begin{figure}[!h]
\centering
\includegraphics[scale=0.3]{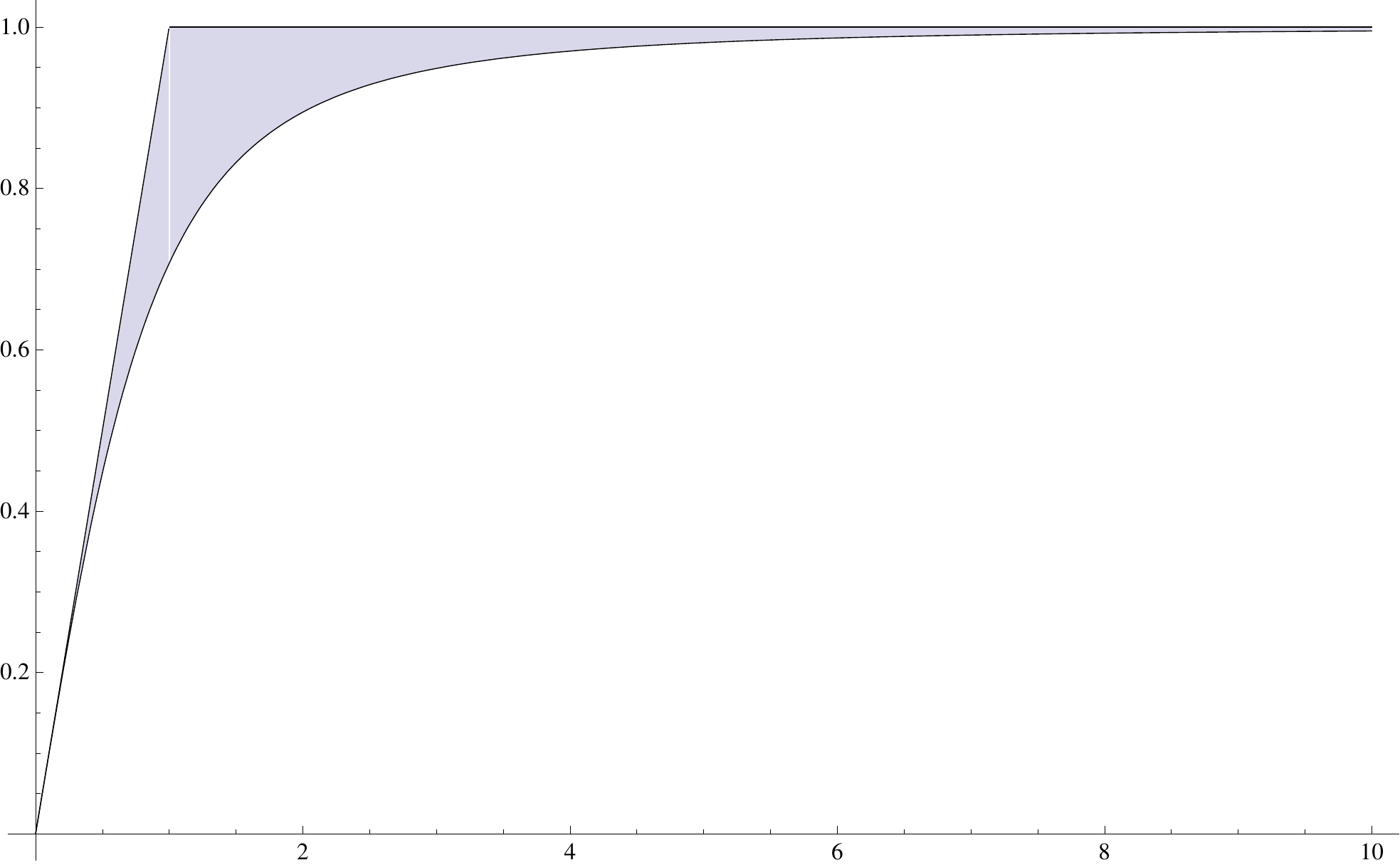}
\caption{\label{fig_approx} Overestimation of the approximation given by \eqq{\ref{approx}}. In this plot we have used $\alpha=1$.}
\end{figure}

As can be seen in figure~\ref{fig_approx}, \eqq{\ref{approx}} overestimate the integrant. Therefore, if the integral converge with this approximation then \eqq{\ref{ISS}} will also converge. The integral reads

\begin{align}
\mathrm{I}_{SS} (\tau)  \cong &
 -\frac{1}{6 \epsilon_{0}c^3\hbar\beta^2 }\int_{0}^{x}{\mathrm{d}x'}\frac{x'}{\sqrt{\alpha^2 + {x'}^2}}\times \Bigg( \mathcal{L}''(x-x') 
  \nonumber\\
&\qquad \qquad  -\frac{1}{\alpha} \Big( \mathcal{L}'(\alpha - x') - \mathcal{L}'(-x') \Big)
\Bigg)\ .
\end{align}

We can make a further approximation. Recall that 
$\alpha^{-1} =\frac{\hbar}{\pi ck_B }\frac{g_0}{T}$
 which generally is much smaller than 1. Indeed we have $\alpha^{-1} \approx 5 \times 10^{-20} \ {g_0}/{T}$ which for nonzero temperature is much smaller than unity. In this manner we can write
\begin{align}
\mathrm{I}_{SS} (\tau) &\cong - \frac{1}{6 \epsilon_{0}c^3\hbar\beta^2 }\int_{0}^{x}\mathcal{L}''(x-x')\frac{x'}{\sqrt{\alpha^2 + {x'}^2}}\mathrm{d}x'\nonumber\\
& = \frac{1}{6 \epsilon_{0}c^3\hbar\beta^2}\left[\mathcal{L}'(x-\alpha) + \mathcal{L}'(0) - \mathcal{L}'(x)\right]\ ,
\end{align}
or explicitly in terms of the proper time
\begin{align}\label{ISSapprox}
\mathrm{I}_{SS} (\tau) &\cong \frac{1}{6 \epsilon_{0}c^3\hbar\beta^2}\left\{\frac{1}{3} - \mathcal{L}' \left[ \frac{\pi }{\hbar\beta}\frac{c}{g_0} \senh \left(\frac{g_0}{c}\tau\right) \right] \right.\nonumber\\
&\qquad \left.+ \mathcal{L}'\left[ \frac{\pi }{\hbar\beta}\frac{c}{g_0} \left(\senh \left(\frac{g_0}{c}\tau\right) -1 \right) \right] \right\}\ .
\end{align}

Then, the  $\mathrm{I}_{SS}$ function is smooth and goes to a constant for $\tau \rightarrow \infty$. Therefore, the behavior of $\mathrm{I}_{CC}$ and $\mathrm{I}_{SS}$ given by eq.'s~(\ref{ICCeval}) and (\ref{ISSapprox}) show that they do not diverge which allow us to neglect the quadratic terms in the nonlocal energy density. To conclude this analysis we need to calculate the two integral in \eqq{\ref{EEnl1}}. They can be treated similarly to $\mathrm{I}_{CC}$ and $\mathrm{I}_{SS}$, \ie
\begin{align}
I_C (\tau)&\equiv \int_{0}^{\tau}\mathcal{C}(\tau, \tau') \cosh \theta' \mathrm{d}\tau' \nonumber\\
&= - \frac{\pi}{6\epsilon_{0}\hbar^2c^3\beta^3 } \mathcal{L}''\left(\frac{c}{g_0}\frac{\pi }{\hbar \beta}\senh \theta\right)
\end{align}
and
\begin{align}
I_S (\tau) &\equiv \int_{0}^{\tau}\mathcal{C}(\tau, \tau') \senh \theta' \mathrm{d}\tau' \nonumber\\
&\cong- \frac{\pi}{6\epsilon_{0}\hbar^2c^3\beta^3}\Big(\mathcal{L}''(x-a) + \mathcal{L}''(x-x) - \mathcal{L}''(x-a)\Big) \nonumber\\
&=0\quad .
\end{align}

Summing all terms, we find that the nonlocal linear correction for the energy density reads
\begin{equation}
\rho_{nl}=
\frac{2\pi}{3}\frac{ g_0 }{\hbar^2c^4 }(k_BT)^3\  \senh \theta  \,  \mathcal{L}''\left[\frac{c}{g_0}\frac{\pi }{\hbar \beta}\senh \theta \right]
\end{equation}

It is convenient to compare the total energy density measured by an accelerated observer with the energy density prescribed by the hypothesis of locality which is given by the ratio
\begin{equation}\label{densNL}
\frac{\rho_{ac}}{\rho'} = 1+\frac{\rho_{nl}}{\rho\left(1+\frac43 \senh^2\theta\right)}\ .
\end{equation}

Note that $\mathcal{L}''(x)$ is a negative function for $x>0$ (see figure~\ref{graficos_L}) showing that, as defined, $\rho_{nl}$ is a negative quantity. Thus, the nonlocal contribution decrease the energy density. In order to estimate the order of magnitude of this effect, it is convenient to recast \eqq{\ref{densNL}} as 
\begin{align}\label{densNL2}
\delta \rho\equiv \frac{\rho_{ac}-\rho'}{\rho'}&=-10\lambda^2\eta^2 \, f(\theta)
\end{align}
where we have defined
\begin{align}
\lambda&=\frac{2\pi \hbar}{c\, k_B}\frac{g_{\oplus}}{T_{cmb}}\ , \label{parameterlambda}\\
\eta&=\frac{1}{2\pi^2}\frac{g_0}{g_{\oplus}}\, \frac{T_{cmb}}{T}\\
x&\equiv \frac{1}{\lambda}\senh \theta\, \label{parameterx}\\
f\left(\theta\right)&\equiv -
\ \frac{\eta x}{1+ \frac43\lambda^2\eta^2\,(\eta x)^2}\ \mathcal{L}''(\eta x)\quad .
\end{align}

The function $f(\theta)$ is positive given that $\mathcal{L}''(x)$ is always negative. In addition, $\mathcal{L}''(x)$ decays faster than its argument and $\eta x.\mathcal{L}''(\eta x)$ is maximum around $\eta x=1$. Therefore, the function $f(\theta)$ is of the order of unit at its maximum. The parameter $\lambda \approx 5,23\times 10^{-19}$ so the nonlocal effects in the energy density is very small and of the order $10^{-36}\, \eta^2$.

\begin{figure}[!h]
\centering
\includegraphics[width=9cm,height=5.3cm]{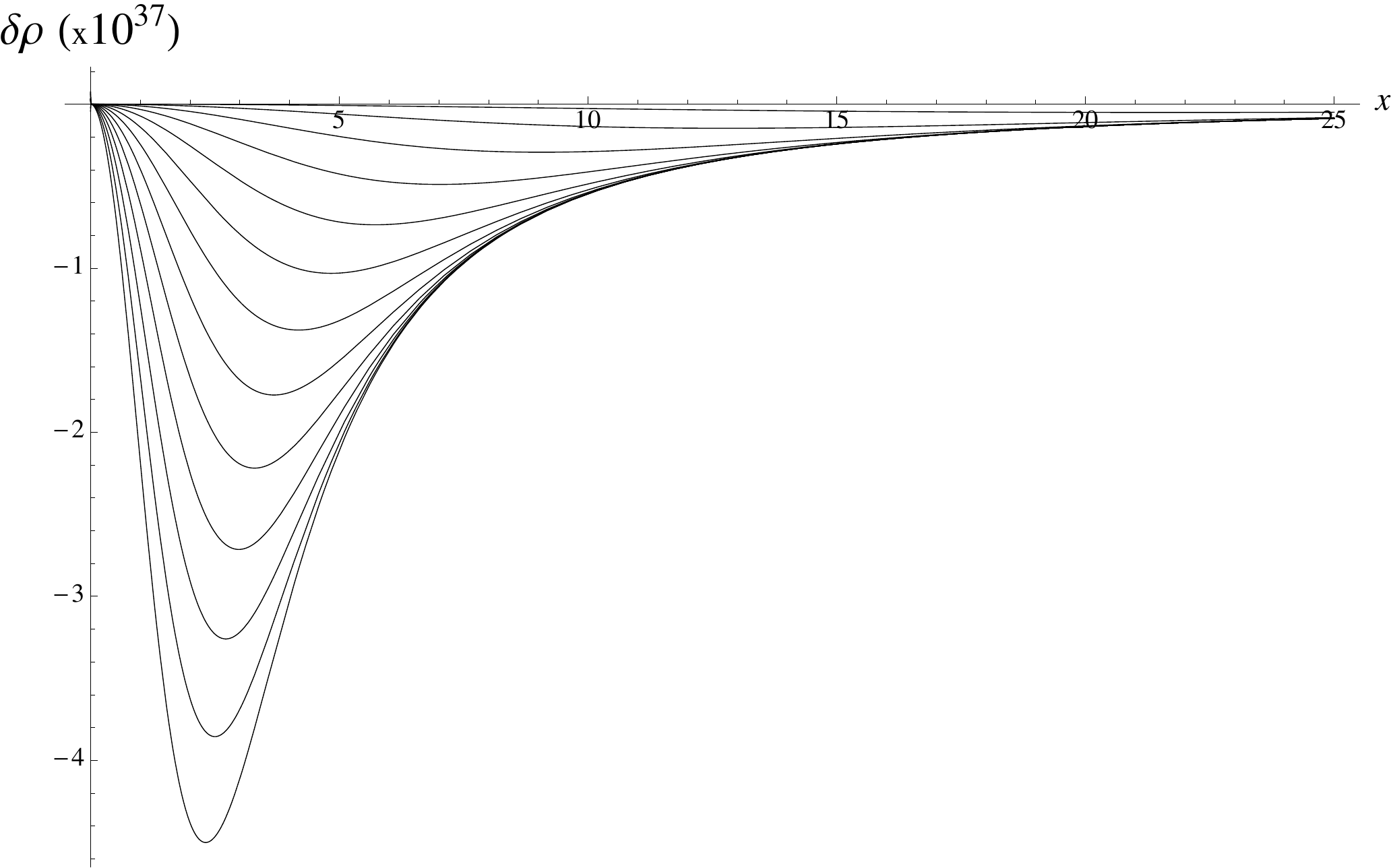}
\caption{\label{fig_density} The nonlocal effect in the energy density. We plot $\delta \rho =({\rho_{ac}-\rho')}/\rho'$ for values of $g_0$ and $T$ that makes $\eta$ varies from $0.1$ to $1$.}
\end{figure}

\subsection{Heat Flux}

The average nonlocal Poynting vector reads
\begin{equation}
\mathcal{S}_i  =  \frac{1}{\mu_0}\epsilon_{ijk}\langle{\mathcal{E}}^j{\mathcal{B}}^k\rangle\quad .
\end{equation}

However, the homogeneity and isotropy conditions impose that any cross term should vanish. The only contributions for the nonlocal Poynting vector comes from $\langle{E_i}^2\rangle$ and $\langle{B_i}^2\rangle$. Thus, the components of the Poynting vector that are perpendicular to the observer's trajectory should vanish inasmuch they contain only cross terms. Indeed, for $  \mathcal{S}_1 $ and $  \mathcal{S}_2  $ there is but terms of the form $\langle  E_{i} E_j \rangle$, $\langle  E_{i} B_j \rangle$ and $\langle  B_{i} B_j \rangle$ with $i\neq j$. All these terms vanish and we find 
\begin{equation}
 \mathcal{S}_1 =  \mathcal{S}_2  =0\quad. 
\end{equation}

For the direction parallel to the observer's trajectory we have
\[
\mathcal{S}_3  =  \frac{1}{\mu_0}\left[\langle{\mathcal{E}}_1{\mathcal{B}}_2\rangle - \langle{\mathcal{E}}_2{\mathcal{B}}_1\rangle\right]=
 \frac{2}{\mu_0}\langle{\mathcal{E}}_1{\mathcal{B}}_2\rangle \quad,
\]
since $\langle{\mathcal{E}}_1{\mathcal{B}}_2\rangle =- \langle{\mathcal{E}}_2{\mathcal{B}}_1\rangle$.
In terms of the background averages we have
\begin{align}
\langle \mathcal{E}_1 \mathcal{B}_2 \rangle =&-{\frac{2\rho}{3c \epsilon_0}}\cosh \theta \senh \theta \nonumber\\
&+\frac{2}{c}\frac{g_{0}}{c}\Big(\cosh \theta \ I_C(\tau)+ \senh \theta \ I_S(\tau)\Big)  \nonumber\\
& -\frac{2}{c}\left(\frac{g_{0}}{c}\right)^2 \mathrm{I}_{SC} (\tau)
\end{align}
where we have defined the integral 
\begin{align}
\mathrm{I}_{SC} (\tau) \equiv &\int_{0}^{\tau} \senh \theta' \ I_C(\tau')\mathrm{d}\tau' \\
=&\int_{0}^{\tau}\int_{0}^{\tau} \mathcal{C}(\tau'',\tau')\senh \theta'\cosh \theta'' \mathrm{d}\tau'\mathrm{d}\tau'' \ .\nonumber
\end{align}

The above integral can be recast as
\begin{equation}
\begin{split}
\mathrm{I}_{SC} (\tau) &= - \frac{1}{6 \epsilon_{0}c^3\hbar\beta^2 }\int_{0}^{x}\int_{0}^{x} \frac{\mathcal{L}'''\left( x'' - x' \right) }{\sqrt{\alpha^2 + {x'}^2}} x'\mathrm{d}x''\mathrm{d}x' \\
&= \frac{1}{6 \epsilon_{0}c^3\hbar\beta^2 }\int_{0}^{x} \frac{\mathcal{L}''(x'-x)-\mathcal{L}''(x')}{\sqrt{\alpha^2 + {x'}^2}}x'\mathrm{d}x'\\
&\cong \frac{1}{6 \epsilon_{0}c^3\hbar\beta^2 } \left\{ \frac{1}{3} - \mathcal{L}'\left[\frac{\pi c}{\hbar \beta g_0}\senh \left(\frac{g_0 \tau}{c}\right)\right] \right\}\ ,
\end{split}
\end{equation}
where the sign $\cong $ refers to an approximation similar to the one used in the $I_{SS}(\tau)$ evaluation. This term is multiplied by a square correction of the observer's acceleration and as before will be neglected. Thus, keeping only first order terms, the non-zero component of the Poynting vector reads
\begin{align*}
 \mathcal{S}_3  = &-\frac{4\rho}{3}\cosh \theta \senh \theta \\
&-\frac{2\pi}{3\hbar^2c^3\beta^3 }\ \frac{g_{0}}{c}\cosh \theta\ 
\mathcal{L}''\left(\frac{c}{g_0}\frac{\pi }{\hbar \beta}\senh \theta \right)\quad ,
\end{align*}
or the ratio between the nonlocal and the local measurement is
\begin{align*}
\frac{\mathcal{S}_3  }{ S'_3  }= &1 +\frac{1}{2\rho}\frac{\pi}{\hbar^2c^3\beta^3 }\ \frac{g_{0}}{c}
\ \frac{\mathcal{L}''\left(\frac{c}{g_0}\frac{\pi }{\hbar \beta}\senh \theta \right)}{\senh \theta}\ ,\\
=&1+\frac{15}{2}\left(\frac{c}{g_0}\frac{\pi }{\hbar \beta}\senh \theta \right)^{-1} 
\mathcal{L}''\left(\frac{c}{g_0}\frac{\pi }{\hbar \beta}\senh \theta \right)\ .
\end{align*}

\begin{figure}[!h]
\centering
\includegraphics[width=8cm]{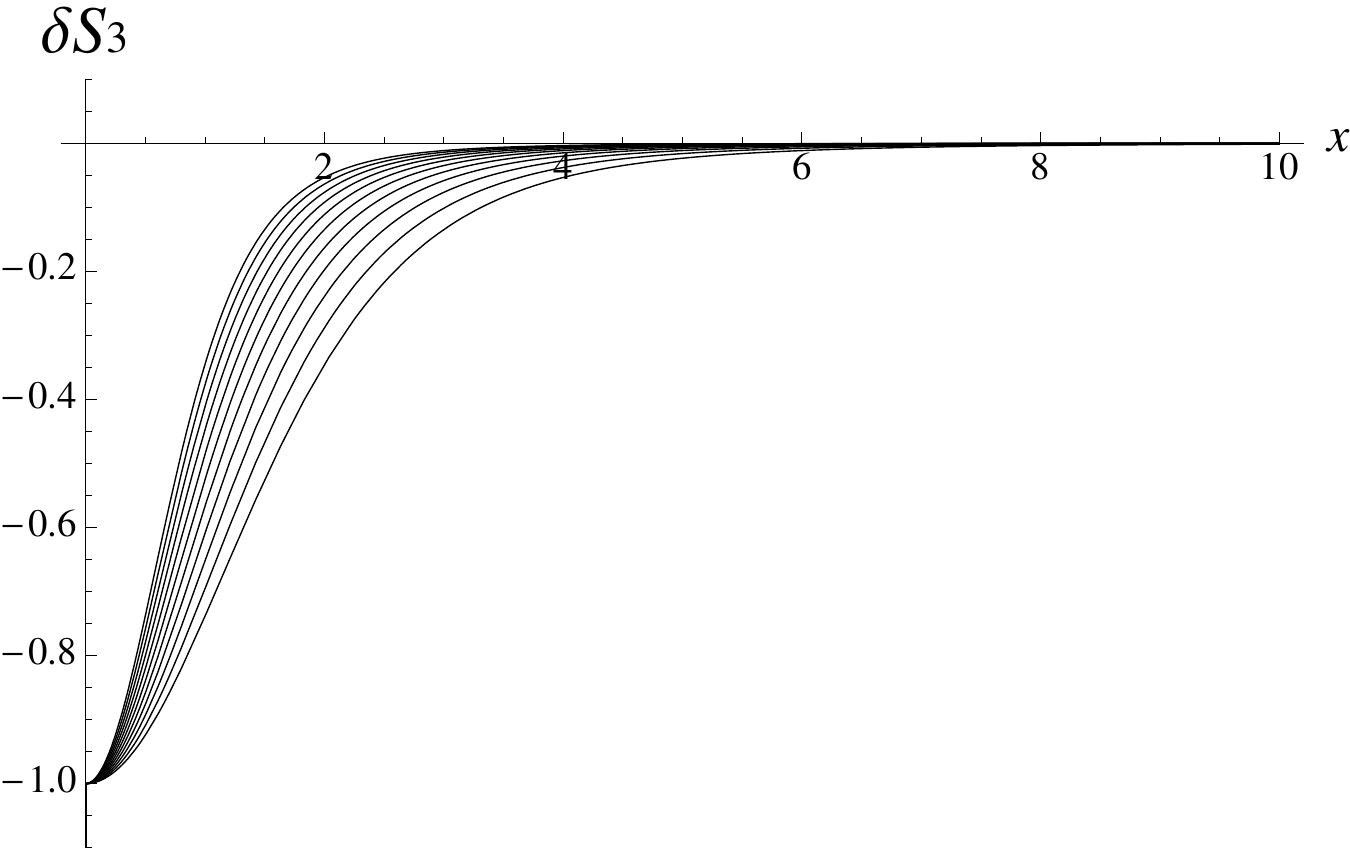}
\caption{\label{S3} The nonlocal effect in the Poynting vector. Here we show the magnitude of the its component parallel to the observer's trajectory. 
We plot $\delta \mathcal{S}_3=\frac{\mathcal{S}_3  - S'_3 }{ S'_3  }$ for values of $g_0$ and $T$ that makes $\eta$ varies from $1$ to $2$.}
\end{figure}

Using the same parameters $\lambda$, $\eta$ and $x$ defined in \eqq{\ref{parameterlambda}}-(\ref{parameterx}), we can rewrite this expression as 
\begin{align*}
\delta \mathcal{S}_3\equiv \frac{\mathcal{S}_3  - S'_3 }{ S'_3  }&=\frac{15}{2}\frac{\mathcal{L}''(\eta x)}{\eta x}
\end{align*}

Again the contribution goes to zero as time evolves and as can be seen by figure~\ref{S3}, the nonlocal effect is significant only at the beginning of the acceleration. The nonlocal effects are of the same order of magnitude that of the local effects, $\delta\mathcal{S}_3$ start at $-1$ and then decays rapidly to zero. However, this effect dies out too fast. The parameter $x$ is inversely proportional to $\lambda \approx 5,23\times 10^{-19}$ , and as soon as $x\sim 2$ the nonlocal effect is already much smaller, $\delta\mathcal{S}_3\sim 0.1$.

\subsection{Maxwell Stress Tensor}

The stress tensor for an accelerated observer can be written as
\begin{equation}
{\mathcal{T}}_{ij} =\epsilon_0 \left[  \frac{1}{2}\left(\mathbf{\mathcal{E}}^2 + c^2 \mathbf{\mathcal{B}}^2\right)\delta_{ij} -{\mathcal{E}}_i{\mathcal{E}}_j - c^2{\mathcal{B}}_i{\mathcal{B}}_j  \right]\  .
\end{equation}

Recalling relations eq.'s (\ref{relacoes1}) and (\ref{relacoes2}), one can immediately check that
\begin{align}
\langle {\mathcal{T}}_{11}\rangle &= \langle {T}_{11}\rangle=\frac13\, \rho\quad ,\\
\langle {\mathcal{T}}_{22}\rangle &= \langle {T}_{22}\rangle=\frac13\, \rho\quad ,
\end{align}
while 
\begin{align}
\langle {\mathcal{T}}_{33}\rangle =& \ \epsilon_0 \Big(2\langle{\mathcal{E}}_1{\mathcal{E}}_1 \rangle - \langle{\mathcal{E}}_3{\mathcal{E}}_3 \rangle\Big) \\
=& \ \frac{1}{3}\, \rho \Big( 1+ 4\senh^2 \theta\Big) \nonumber\\
&+\frac{2\pi}{3\hbar^2c^3\beta^3}\left( \frac{g_0}{c}\right) \senh \theta  \, \mathcal{L}''\left(\frac{c}{g_0}\frac{\pi }{\hbar \beta}\senh \theta \right)\ .\nonumber
\end{align}

All cross terms like $\langle {\mathcal{T}}_{12}\rangle$ or $\langle {\mathcal{T}}_{23}\rangle$ are zero. Thus, the ratio of the nonlocal contribution to the local measurement of the nonzero component of the stress tensor read
\begin{align}
\frac{\langle {\mathcal{T}}_{11}\rangle }{\langle {T'}_{11}\rangle}&=\frac{\langle {\mathcal{T}}_{22}\rangle }{\langle {T'}_{22}\rangle}=1\\
\frac{\langle {\mathcal{T}}_{33}\rangle }{\langle {T'}_{33}\rangle}&=1+ 30\frac{g_0 \hbar \beta}{\pi c}\frac{\sinh \theta}{1+4\sinh^2 \theta}\, \mathcal{L}''\left(\frac{c}{g_0}\frac{\pi }{\hbar \beta}\senh \theta \right)
\end{align}

The fractional nonlocal effect for the stress tensor can be defined as
\begin{equation}
\delta T_{33}=- 30\lambda^2\eta^2\  j(\theta)\, ,
\end{equation} 

where we have defined 
\[ 
j(\theta)\equiv -
\frac{\eta x}{1+4\lambda^2 \eta^2(\eta x)^2}  \, \mathcal{L}''(\eta x)\ .
\]

Similarly to $f(\theta)$ defined for the fractional nonlocal energy density, the function $j(\theta)$ is positive and at most of order 1. Thus, the nonlocal effects scales with $\lambda^2$, \ie this effects is of order of $10^{-35}\, \eta^2$.

\begin{figure}[!h]
\centering
\includegraphics[scale=0.4]{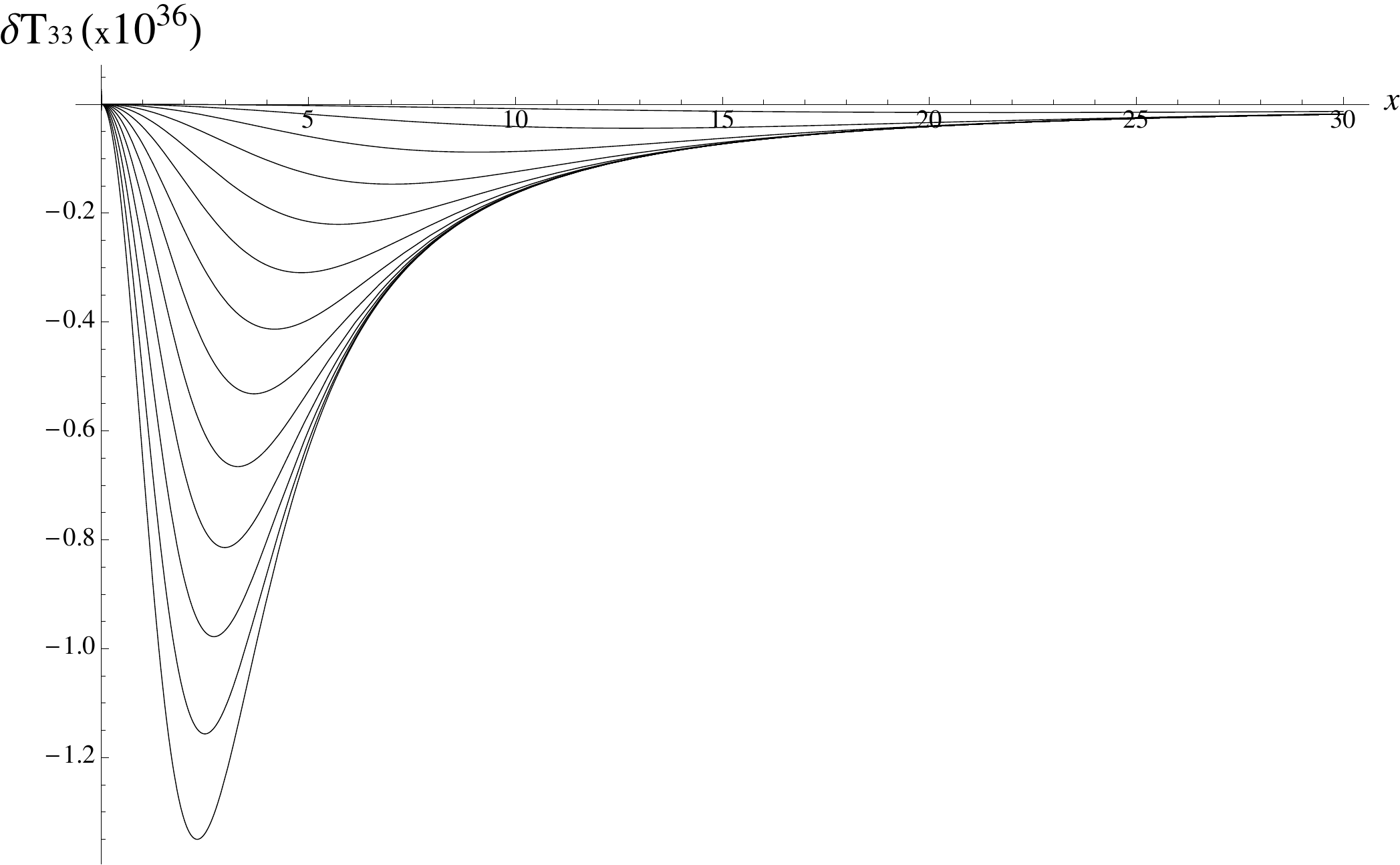}\label{pressao}
\caption{Nonlocal change in the $\mathcal{T}_{33}$ component of the stress tensor for an observer world line with linear acceleration in this same direction. 
We plot $\delta \mathcal{T}_{33}=\frac{\mathcal{T}_{33}  - T'_{33} }{ T'_{33}  }$ for values of $g_0$ and $T$ that makes $\eta $ varies from $.1$ to $1$.}
\end{figure}

The above nonlocal correction represents a change in the measured pressure of the background radiation. The transverse direction with respect to the observer's trajectory do not change while the pressure along the observer's path is suppressed by the nonlocal effect. Analogously to the heat flux and the energy density, the Maxwell's tensor decrease due to nonlocal effects. \\

\section{Conclusion}
In this paper we applied the nonlocal formalism for accelerated observers, developed by Mashhoon and collaborators, to analyze how it modifies the thermodynamics properties of an electromagnetic radiation field. In particular, we studied the case of a homogeneous and isotropic blackbody radiation using an average over ensemble to define the space average fields. Considering a linear accelerated observer, we disentangled the pure nonlocal effects from the common relativistic effects. In this case, the coherence function shows that the nonlocal effects are all transient and quickly decays. Thus, at least for the specify example studied in this paper, we do not except to identify any measurable effect. Notwithstanding, there is no reason for this quickly transient decaying to be a general behavior. For circularly accelerated observer, the nonlocal effects might leave some measurable imprint in the black body radiation.

\section*{ACKNOWLEDGEMENTS}

We would like to thank CAPES and CNPq of Brazil for financial support.


\end{document}